\documentclass[prb,aps,amsmath, twocolumn, amssymb]{revtex4}
\usepackage{graphicx}
\pdfoutput=1
\usepackage{amsmath}
\usepackage{color}
\usepackage{float}
\usepackage{tabularx}
\newcolumntype{L}[1]{>{\raggedright\arraybackslash}p{#1}}
\newcolumntype{C}[1]{>{\centering\arraybackslash}p{#1}}
\newcolumntype{R}[1]{>{\raggedleft\arraybackslash}p{#1}}

\begin{document}

\title{High spin polarization and the origin of unique ferromagnetic ground state in CuFeSb}

\author{Anshu Sirohi$^1$}

\author{Chandan K. Singh$^2$}

\author{Gohil S. Thakur$^3$}

\author{Preetha Saha$^1$}

\author{Sirshendu Gayen$^1$}

\author{Abhishek Gaurav$^1$}

\author{Shubhra Jyotsna$^1$}

\author{Zeba Haque$^3$}

\author{L. C. Gupta$^3$}

\author{Mukul Kabir$^2$}

\author{Ashok K. Ganguli$^{3,4}$}

\author{Goutam Sheet$^1$}

\email{goutam@iisermohali.ac.in}

\affiliation{$^1$Department of Physical Sciences,  
Indian Institute of Science Education and Research Mohali,
Sector 81, S. A. S. Nagar, Manauli PO 140306, India}

\affiliation{$^2$Department of Physics, Indian Institute of Science Education and Research, Pune 411008, India}

\affiliation{$^3$Department of Chemistry, Indian Institute of Technology, New Delhi
110016, India}

\affiliation{$^4$Institute of Nano Science \& Technology, Mohali 160064, India}

\begin{abstract}

CuFeSb is isostructural to the ferro-pnictide and chalcogenide superconductors and it is one of the few materials in the family that are known to stabilize in a ferromagnetic ground state. Majority of the members of this family are either superconductors or antiferromagnets. Therefore, CuFeSb may be used as an ideal source of spin polarized current in spin-transport devices involving pnictide and the chalcogenide superconductors. However, for that the Fermi surface of CuFeSb needs to be sufficiently spin polarized. In this paper we report direct measurement of transport spin polarization in CuFeSb by spin-resolved Andreev reflection spectroscopy. From a number of measurements using multiple superconducting tips we found that the intrinsic transport spin polarization in CuFeSb is high ($\sim$ 47\%). In order to understand the unique ground state of CuFeSb and the origin of large spin polarization at the Fermi level, we have evaluated the spin-polarized band structure of CuFeSb through first principles calculations. Apart from supporting the observed 47\% transport spin polarization, such calculations also indicate that the Sb-Fe-Sb angles and the height of Sb from the Fe plane is strikingly different for CuFeSb than the equivalent parameters in other members of the same family thereby explaining the origin of the unique ground state of CuFeSb.  

\end{abstract}

\maketitle

It is believed that the exotic superconductivity in the iron-based pnictide and chalcogenide superconductors originates from a magnetically driven pairing mechanism where the superconducting order is thought to be coupled with spin fluctuations\cite{1-Ahilan, 2-Nakai, 3-Christianson, 4-Lumsden, 5-Chi}. Therefore, in order to understand the nature of coupling between the superconducting order and spin fluctuations, it is imperative to understand the nature of magnetism and the spin fluctuation in the parent compounds from which the superconducting states are derived through doping. Usually the parent compounds of the pnictide and the chalcogenide superconductors are known to be antiferromagnetic where the antiferromagnetism is thought to be promoted by spin density waves (SDW) associated with Fermi surface nesting\cite{6-Cruz, 7-Huang, 8-Bao, 9-Dong,10-Singh, 11-Ma, 12-Cvetkovic, 13-Mazin}. More recently it has been shown that CuFeSb, which is isostructural to the iron-based layered superconductors e.g., Li(Na)FeAs\cite{Arsenic}, has a ferromagnetic ground state\cite{14-Qian}. A close relative of this compound CuFeAs stabilizes in an antiferromagnetic ground state with a Neel temperature of 9 K\cite{Gohil}. The ferromagnetic order in CuFeSb is thought to originate from the large height of Sb from the Fe plane. This fact also supports the hypothesis that the competing magnetic interactions in ferropnictide superconductors is decided by the anion height i.e., there is a gradual change in the magnetic properties from superconductivity to
antiferromagnetism to ferromagnetism on moving in the increasing order of anion height from LiFeAs to CuFeAs to CuFeSb \cite{14-Qian, 14a-Moon, Gohil}.  CuFeSb is known to be one of the very few materials in the FeAs or FeSb family that shows a ferromagnetic ground state. Therefore it is most important to understand the Fermi surface properties of this unique system by spectroscopic measurements -- particularly, the nature of the Fermi surface spin polarization and the degree of spin fluctuations. Furthermore, owing to the structural similarities between CuFeSb and the pnictide and chalcogenide superconductors, a highly spin-polarized Fermi surface of CuFeSb might be useful as a source of spin-polarized current in devices involving the pnictide and the chalcogenide superconductors for spin transport related experiments.

In this paper we have employed spin-resolved Andreev reflection spectroscopy using conventional superconducting tips to measure the spin polarization at the Fermi level of CuFeSb \cite{15-Soulen}. From the analysis of the Andreev reflection data between the superconductor and the ferromagnet, we found the evidence of a high degree of transport spin polarization approaching $47\%$ in CuFeSb. In addition to the insight regarding the magnetic interactions in the ferropnictide superconductors that might emerge from this result, it should be noted that the ferromagnets with transport spin polarization approaching $50\%$ are potential candidates as spin source in spintronic devices.
\begin{figure}
\begin{center}
\centering
\includegraphics[scale=0.365]{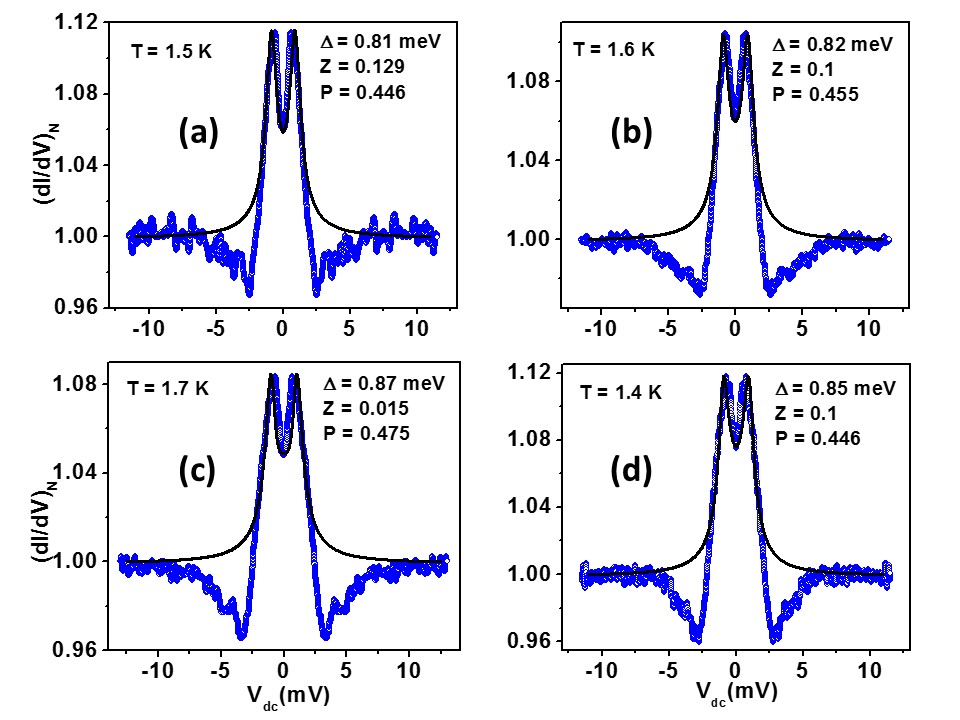}
\end{center}
\caption {Representative point-contact Andreev reflection spectra between superconducting Nb tip and CuFeSb. The blue dotted lines show the raw experimental data and the solid black lines show the theoretical fit (see text). All the spectra are normalized with respect to the high-bias conductance. $(dI/dV)_N$ indicates normalized differential conductance. The measurement temperature and the fitting parameters are listed in the respective panels.} 
\end{figure}

The transport through a ballistic point-contact between a normal metal and a superconductor is dominated by Andreev reflection that involves the reflection of a spin up (down) electron as a spin down (up) hole from the interface\cite{16-Tinkham}.  The Andreev reflection spectra ($dI/dV$ vs. $V$) are traditionally analyzed by the theory developed by Blonder, Tinkham and Klapwijk (BTK)\cite{17-BTK}. This theory assumes a delta-function potential barrier whose strength is characterized by a dimensionless parameter $Z$ which is proportional to the strength of the barrier. Ideally, for elemental superconductors where the quasi-particle life-time is very large, the Andreev reflection spectra can be fitted using two fitting parameters, $Z$ and $\Delta$, the superconducting energy gap. For all non-zero values of $Z$, the $dI/dV$ spectra show a double-peak structure symmetric about $V = 0$. However, in the system where the life-time of the quasi-particles is finite due to inelastic processes at the interface, the Andreev reflection spectrum undergoes broadening. This broadening is accounted for in the modified BTK theory where a complex component $i\Gamma$ is artificially added to the quasi-particle energy ($E$)\cite{18-BSCCO}. In such cases the spectra are analyzed using three fitting parameters, $Z$, $\Delta$ and $\Gamma$.
\begin{figure}
\begin{center}
\centering
\includegraphics[scale=0.365]{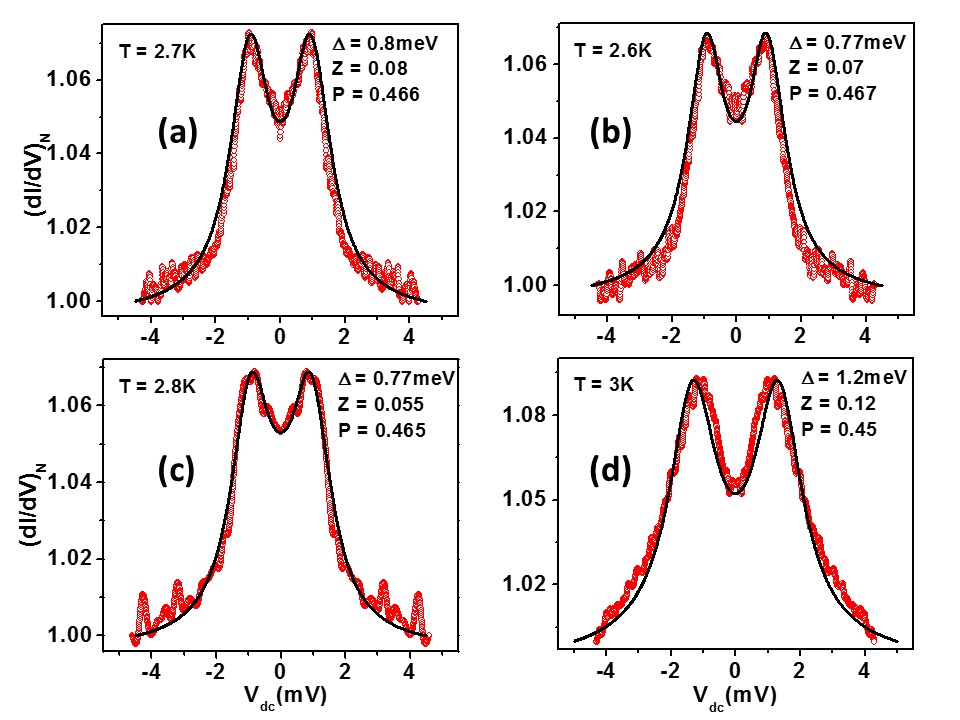}
\end{center}
\caption { Representative point-contact Andreev reflection spectra between superconducting Pb tip and CuFeSb. The red dotted lines show the raw experimental data and the solid black lines show the theoretical fit (see text).  All the spectra are normalized with respect to the high-bias conductance. $(dI/dV)_N$ indicates normalized differential conductance. The measurement temperature and the fitting parameters are listed in the respective panels.} 
\end{figure}

When the metal in the metal-superconductor point-contact is a ferromagnet, the Fermi level is expected to be spin polarized\cite{13-Mazin}. This means, the density of states of the up-spin electrons ($N_\uparrow$) is not equal to the density of states of the spin down electrons ($N_\downarrow$) at the Fermi level. Therefore,  $|N_\uparrow - N_\downarrow|$ electrons encountering the interface cannot undergo Andreev reflection because they do not find accessible states in the opposite spin band. Therefore, in a point-contact between a ferromagnetic metal and a conventional superconductor, Andreev reflection is suppressed. By measuring the degree of this suppression the spin-polarization of the Fermi surface is measured\cite{19-Pratap, 20-Goutam, 21-Surjeet}. In order to extract the absolute value of the Fermi level spin polarization, first the BTK current is calculated for zero spin polarization ($I_{BTKu}$) and  $100\%$ spin polarization ($I_{BTKp}$) respectively. Then the current for an intermediate spin polarization $P_t$ is calculated by interpolation between ($I_{BTKu}$) and ($I_{BTKp}$) following the relation $I_{total} = I_{BTKu}(1-P_t) + P_t I_{BTKp}$. The derivative of $I_{total}$ with respect to $V$ gives the modified Andreev reflection spectrum with finite spin polarization in the metal. This model is used to analyze the spin-polarized Andreev reflection spectra obtained between a ferromagnetic metal and a superconductor by using four fitting parameters $Z$, $\Delta$ and $\Gamma$ and $P_t$. It should be noted that for the measurement of the spin polarization of ferromagnets usually standard conventional superconducting probes are used for which the value of $\Delta$ is known. In addition, in order to have superconductivity, $\Gamma$ cannot be arbitrarily large with respect to $\Delta$. Therefore, effectively only two parameters, $Z$ and $P_t$, are tuned freely during the analysis of spin-polarized Andreev reflection spectra. \\

\begin{figure}
\begin{center}
\centering
\includegraphics[scale=0.60]{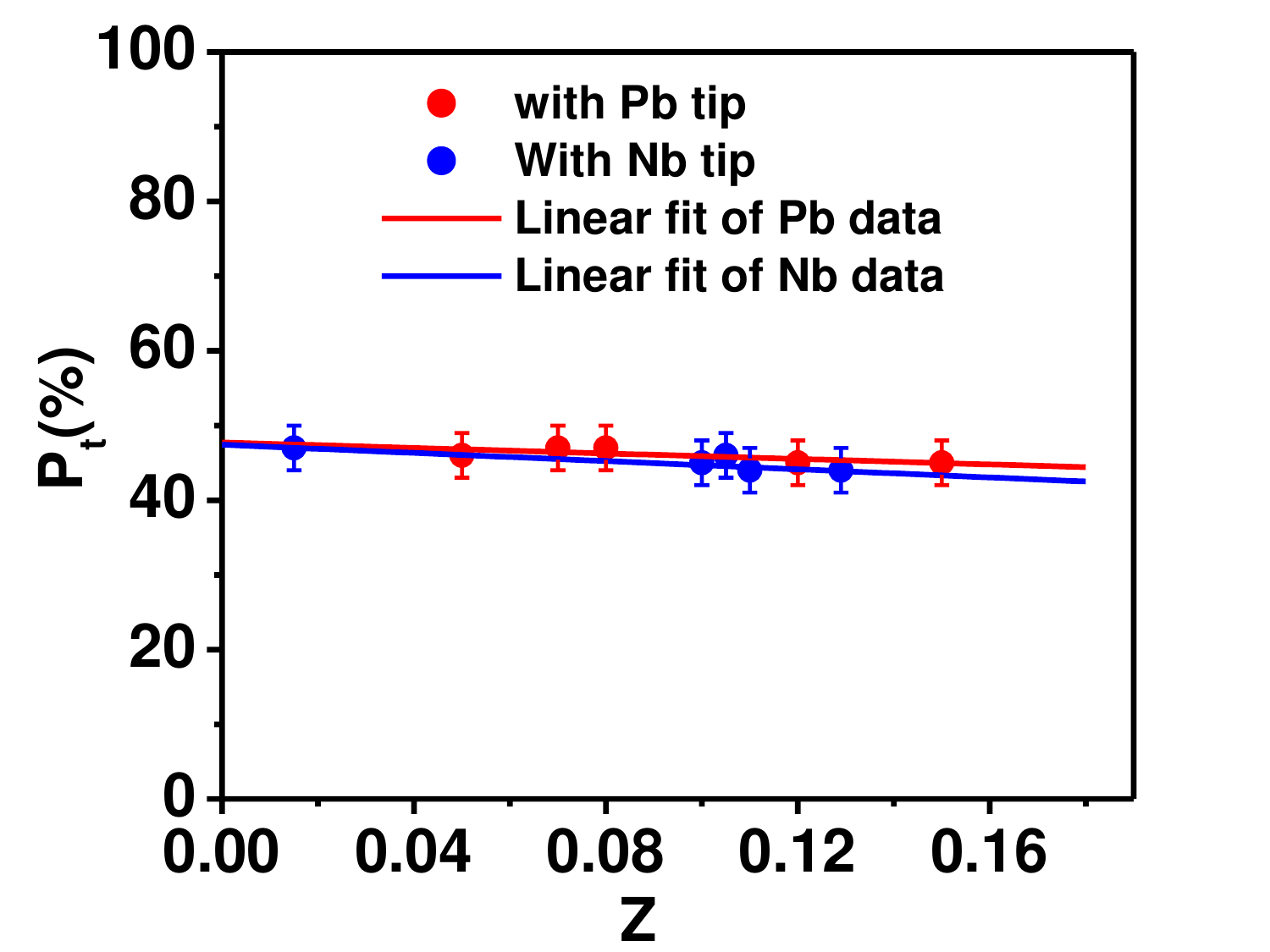}
\end{center}
\caption { Dependence of transport spin-polarization on Z for point-contacts on CuFeSb with Nb and Pb tips. The red circles show the data points for Nb tips and the blue stars show the data points for Pb tips. The solid lines show the respective linear fits: Red line for Nb tips and blue line for Pb tips. Though $P_t$ shows different dependence on $Z$, the intrinsic value of $P_t$ extracted for $Z = 0$ is identical for both Nb and Pb tips. } 
\end{figure}

All the measurements reported in this paper were performed on a polycrystalline pellet of CuFeSb. CuFeSb shows a ferromagnetic transition around 380 K\cite{14-Qian}.                                                                                                                                                                                                  
The Andreev reflection spectroscopic measurements were performed by measuring the transport characteristics of several ballistic point-contacts between CuFeSb and the elemental superconductors niobium (Nb) and lead (Pb) respectively using a home-built point-contact spectroscopy probe in a liquid helium cryostat. First the polycrystalline pellet of CuFeSb was polished and mounted on a copper disc which is used as the sample holder in the home-built probe. A calibrated cernox thermometer and a heater were attached to the same copper disc for precise measurement of the temperature and for controlling the sample temperature during the Andreev reflection spectroscopic measurements. Two 100 $\mu m$ thick gold wires were mounted on the sample for transport measurements. The superconducting tips were fabricated from 250 $\mu m$  diameter wires of Nb and Pb respectively. The tips were mounted on a teflon piece connected to the head of a 100 threads per inch differential screw. Two more 100 $\mu m$ thick gold wires were mounted on the  tip. The probe was then mounted inside the static variable temperature insert (VTI) of a liquid-helium cryostat. The static VTI was surrounded by a dynamic VTI with a micro-capillary that allowed us to perform measurements down to 1.4 K.

The ballistic point-contacts between the sample and the tips were fabricated and controlled by moving the tip up and down by rotating the differential screw manually. For electrical measurements a lock-in modulation technique was employed where a dc current was coupled with a small current drawn from the sinusoidal output of a digital lock-in amplifier (Stanford Research Systems, model: SR-830) through a passive current source. The modulated current was sent through the point-contacts. The dc component of the voltage drop across the point-contacts was measured by a digital voltmeter (Keithley Model: 2000) and the ac component was measured by the same lock-in amplifier. The data acquisition was done by using a lab-view programme developed in house. The ac voltage drop measured in the lock-in is proportional to differential change in voltage ($dV$).  $dI/dV$ plotted against the dc voltage drop $V$ across a point-contact gives the point-contact Andreev reflection spectrum. $dI$ is proportional to the excitation current set during the experiment.

In Figure 1 (a,b,c,d) we show four representative Andreev reflection spectra between a Nb-tip and CuFeSb. The spectra clearly show the double-peak structure symmetric about $V = 0$, which is the hallmark of Andreev reflection. For low values of $Z$, these peaks appear close to the energy gap of the superconductor. The solid lines show the theoretical fits as per the model described above. The superconducting energy gap of niobium is found to vary from contact to contact and fall approximately between $0.8 meV$ and $0.9 meV$ for all the point-contacts that we have analyzed, indicating that the proximity of the ferromagnet suppresses the superconductivity of the point-contacts slightly. The value of $\Gamma$ remained zero for all the spectra, which means the broadening due to finite quasi-particle lifetime is absent at the point-contact. This fact also indicates that the spin-fluctuation in the system is not significant as strong spin-fluctuations is also known to give rise to large $\Gamma$\cite{Sourin}. Therefore, the theoretical fits are obtained by essentially tuning two parameters namely $Z$ and $P_t$, this makes the fit accurate and the fitting parameters unique. It is found that the raw data deviate slightly  from the fit at certain points (notice the dip structures in $dI/dV$). Such deviation is known to originate from the critical current of the superconductor when a small part of the Maxwell's resistance is also measured along with the Sharvin resistance in the point-contacts close to the ballistic regime.\cite{22-Goutam} The experiments were repeated with superconducting Pb-tips to confirm the reproducibility (Figure 2). It is seen that the theoretical fitting to the experimentally obtained spectra obtained with Pb tips is very good and the critical current driven dips are absent. This is due to the existence of a thick oxide layer on the surface of Pb that is broken by the application of mechanical pressure for the formation of point contacts thereby leading to extremely small diameter contacts where ballistic transport dominates. The value of $\Gamma$ remained almost zero for all the fittings and the superconducting energy gap was found to vary approximately between $0.8 meV$ and $1.2 meV$ which is consistent with the superconducting energy gap of bulk Pb.

\begin{table*}
 \caption{Refined structural parameters for CuFeSb obtained at 300K, which is below the ferromagnetic transition temperature. The number in the square bracket indicates the number of symmetric bond lengths. Calculated structural parameters within DFT-PBE formalism are in excellent agreement with the present and previous experimental measurements.~\citep{PhysRevB.85.144427}}
 \label{table1}
 \begin{tabular}{L{3.2cm}L{1.5cm}C{1.5cm}C{1.5cm}C{1.5cm}C{1.5cm}C{2.7cm}}
 \hline
 \hline \\[-1ex]
 &$P4/nmm$: & \multicolumn{5}{l}{ $a=b=$ 3.9117(3) \AA, $c=$ 6.2619(4) \AA \ \ (300K)} \\[0.7ex] 
  & & \multicolumn{5}{l}{$a=b=$ 3.9347(2) \AA, $c=$ 6.2515(4) \AA \ \ (400K, Ref. [\citep{PhysRevB.85.144427}])} \\[0.7ex] 
 & & \multicolumn{5}{l}{$a=b=$ 3.91 \AA, $c=$ 6.32 \AA \ \ (DFT-PBE)} \\[1ex]
 \hline \\[-1ex]
  &  & $x$ & $y$ & $z$  & $B$ (\AA$^{2}$) & Wyckoff positions \\[0.7ex] 
  \hline \\
  Atomic coordicates & Cu & 0.25 & 0.25 & 0.7127(3)             & 1.68(6) & 2c \\
                                        &  Fe & 0.75 & 0.25  & 0                  & 1.69(1) & 2b \\ 
                                         &  Sb & 0.25 & 0.25  & 0.2981(1)    & 1.19(4) & 2c \\ [1.3ex] 
                                         
                                         & \multicolumn{4}{c}{Bond lengths (\AA)} & \multicolumn{2}{c}{Sb--Fe--Sb angles (${^{\circ}}$)} \\
  &   Fe--Sb [4] & Fe--Fe [4] & Cu--Sb [1] & Cu--Sb [4] &  $\alpha$ & $\beta$\\[0.2ex] 
\cline{2-7}\\
Experiment         &  2.704(1)   & 2.766(1) & 2.596(3) & 2.767(2) & 92.68(2) & 118.47(2) \\
Ref. [\citep{PhysRevB.85.144427}]                &  2.693(1)   & 2.782(2) & 2.660(3) & 2.784(2) & 93.86(5) & 117.80(3)\\ 
DFT-PBE              &  2.72           &  2.77          & 2.69         & 2.77 & 91.98 & 118.86\\  [1ex]
 \hline
 \hline
 \end{tabular}
 \end{table*}

\begin{figure*}
\centering
\includegraphics[scale=0.5]{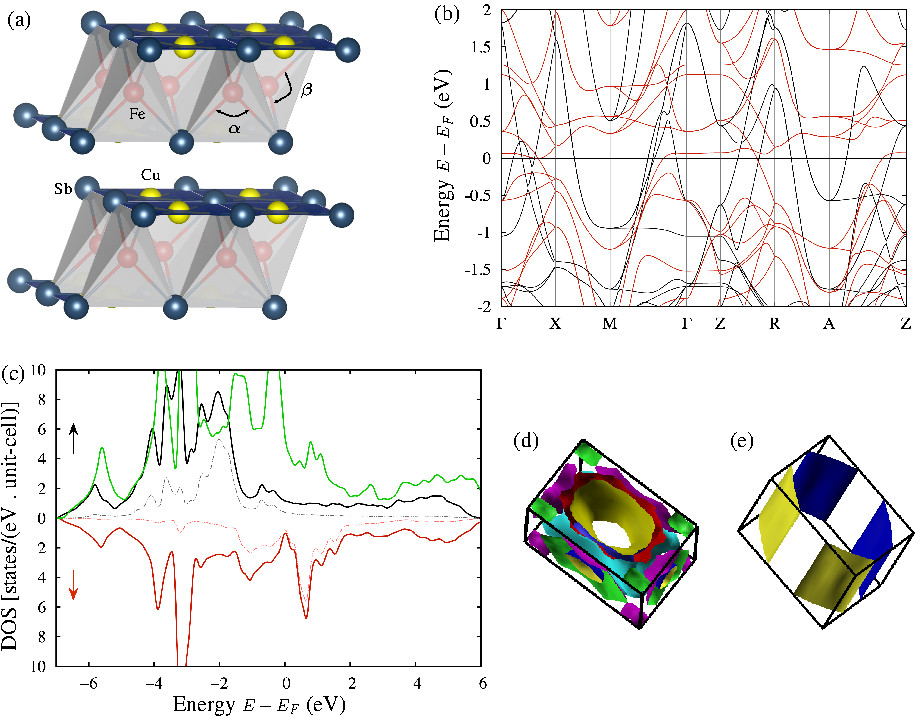}
\caption{(a) Crystal structure of CuFeSb, which has tetragonal $P4/nmm$ symmetry. Different Sb--Fe--Sb angles, $\alpha$, and $\beta$ are indicated. (b) Calculated spin-polarized band structure within density functional theory indicates CuFeSb to be metallic. Bands corresponding to the up (down) channel are indicated in black (red) color.  (c) The nonmagnetic (green line) and ferromagnetic DOS for CuFeSb.  For nonmagnetic DOS, the conduction band is very narrow, which is entirely of Fe-$d$ character, and  the number of states at the Fermi level is very high, resulting in ferromagnetic Fe-layer. The dashed lines for the spin-polarized DOS indicate the Fe-$d$ character for up (black) and down (red) spin electrons.  Spin-polarized Fermi surfaces are shown in (d) for up and (e) for down spin electrons.}
\label{figure-theory}
\end{figure*}

The dependence of $P_t$ on $Z$ for both Nb and Pb based point-contacts is shown in Figure 3. For most of the Nb/CuFeSb point-contacts, the value of $Z$ was found to be small ($< 0.2$) For such point-contacts, the maximum measured value of $P_t $ is found to be $47\%$. For both the Nb/CuFeSb and the Pb/CuFeSb point contacts $P_t$ did not change noticeably with $Z$. It should be noted that within the BTK formalism, in fact, no correlation between $P_t$ and $Z$ is expected. A close inspection however reveals that $P_t$ shows slight dependence on $Z$ and the dependence is linear -- the measured value of $P_t$ decreases slightly with increasing strength of the barrier (represented by $Z$).  Although this dependence is not understood within the BTK formalism, for a vast majority of ferromagnetic point-contacts such dependence was found in the past. In such cases the dependence is attributed to spin depolarization at a magnetically disordered scattering barrier formed at the interface.  In such cases, the conventional way of finding the intrinsic transport spin polarization is to extrapolate the $P_t$ vs. $Z$ curve to $Z = 0$. By doing this extrapolation, the intrinsic $P_t$ is found to be approximately $47\%$ which is nearly equal to the value measured with the Nb tip. Therefore, it is rational to conclude that the degree of spin polarization at the Fermi level of CuFeSb is approximately $47\%$.

In order to understand the unique ferromagnetic ground state of CuFeSb and the origin of the observed high spin polarization, we have performed spin-polarized band structure calculation of CuFeSb. The first-principles calculations were performed within spin-polarized density functional theory (DFT)~\citep{PhysRevB.47.558, PhysRevB.54.11169} using projector augmented wave formalism~\citep{PhysRevB.50.17953} with kinetic-energy cut-off of 370 eV for the wave functions. The exchange-correlation is described with Perdew-Burke-Ernzerhof (PBE) form of generalized gradient approximation,~\citep{PhysRevLett.77.3865} and the reciprocal space integration was carried out using 31$\times$31$\times$17 Monkhorst-Pack k-point sampling. Atomic positions as well as the volume and shape of the unit cell were relaxed until the force components on each atom are less than 0.01 eV/\AA. The structural parameters of the optimized tetragonal [$P4/nmm$, shown in Fig.~\ref{figure-theory}(a)] CuFeSb are in excellent agreement with the current and previous~\citep{PhysRevB.85.144427} experimental measurements (Table~\ref{table1}).  The CuFeSb crystal structure is analogous to other iron pnictides and chalcogenides, though there are some striking differences from the superconducting ones.~\citep{B813153H} For example, while compared with the superconducting LiFeAs, (a) the Sb--Fe--Sb angles are substantially different than As--Fe--As angles, and (b) the Fe--Sb bond length is significantly larger (Table~\ref{table1}). Thus, the height of Sb (As) from the Fe plane is strikingly different for CuFeSb [$z_{\rm Sb}$=1.89 \AA~ (DFT-PBE), and 1.87 \AA~ (experiment)] and LiFeAs ($z_{\rm As}$=1.51 \AA),~\citep{B813153H} and this $z_{\rm anion}$ is predicted to play an important role in determining the electronic structure.~\citep{PhysRevLett.104.057003} 

Although CuFeSb is isostructural to other iron pnictides and chalcogenides, the electronic and magnetic structures are calculated to be completely different due to profound difference in structural parameters discussed above. Large class of  iron pnictides and chalcogenides are found to be antiferromagnetic and superconducting in ambient or  pressured or chemically modified environment.~\citep{Paglione-Nature645, Basov-nature272} In remarkable contrast, we find CuFeSb to be a ferromagnetic metal. Calculated band structure and density of sates (DOS) are shown in Fig.~\ref{figure-theory}(b) and (c), which indicate CuFeSb to be metallic. Present theoretical results are in good agreement with earlier magnetization and neutron scattering measurements, with 375 K transition temperature.~\citep{PhysRevB.85.144427} Calculated Fe moment ($m_{\rm Fe}=$2.4 $\mu_B$) is larger than the experimental moment (1.85 $\mu_B$) at 20K,which further increases with decreasing temperature.~\citep{PhysRevB.85.144427}   However, the calculated $m_{\rm Fe}$ is comparable to the earlier theoretical prediction for FeSeTe (2.94 $\mu_B$ with $z_{\rm Te}$=1.81 \AA).~\citep{PhysRevLett.104.057003} The nonmagnetic DOS in Fig.~\ref{figure-theory}(c) reveal that the conduction electrons at the Fermi level are of Fe-$d$ character, and belong to a narrow band with large density of states, and due to this Stoner instability the Fe layer becomes ferromagnetic.  It would be interesting to compare the DOS at the Fermi level $N(E_F)$ for CuFeSb, which is metallic ferromagnet with the superconducting counterparts. The nonmagnetic $N(E_F)$ is found to be 5.6 eV$^{-1}$ per unit cell for CuFeSb [Fig.~\ref{figure-theory}(c)], which is significantly larger than superconducting LiFeAs (4.5 eV$^{-1}$), and other Fe-based superconductors.~\citep{PhysRevLett.100.237003, PhysRevLett.101.207004, PhysRevB.85.094510}  In contrast, $N(E_F)$ is comparable to MgFeGe which is isoelectronic to LiFeAs but non-superconducting similar to CuFeSb.~\cite{doi:10.7566/JPSJ.82.034714} Large number of states at the Fermi level and in the vicinity is conventionally connected to the number of condensed Cooper pairs, which enhance superconductivity. However, for CuFeSb the narrow $d$ band near the Fermi level and very high $N(E_F)$ the magnetic Stoner instability takes over, and the Fe-layer becomes ferromagnetic. This is similar to the case in ferromagnetic MgFeGe.~\cite{doi:10.7566/JPSJ.82.034714}

Next we calculate the transport spin polarization $P_t$, which is in excellent agreement with the present spin-resolved Andreev reflection spectroscopy measurement.  Within classical Bloch-Boltzmann transport theory, $P_t$ can be defined in terms of spin-dependent current densities, and in general,~\citep{PhysRevLett.83.1427}
$$
P_t^n = \frac{\langle N(E_F) v_F^n \rangle_{\uparrow} - \langle N(E_F) v_F^n \rangle_{\downarrow}}{\langle N(E_F) v_F^n \rangle_{\uparrow} + \langle N(E_F) v_F^n \rangle_{\downarrow}},
$$
where $v_F$  is the spin polarized Fermi velocity of electrons. Thus by definition $P_t^n$ is connected to the the spin polarization measured in various experiments, spin resolved photoemission ($n=0$) and point-contact spectroscopy in ballistic ($n=1$) and diffusive ($n=2$) regimes.  In the present experimental setup, $P_t$ has been measured in the ballistic regime.   
We calculated the spin-polarized DOS, which is shown in Fig.~\ref{figure-theory}(c), along with the spin-polarized Fermi surface [Fig.~\ref{figure-theory}(d) and (e)]. Similar to the non-magnetic case, the conduction electrons at the Fermi level have Fe-$d$ character. Further, $N_{\uparrow}(E_F)$ is found to be 12.6\% larger than $N_{\downarrow}(E_F)$ for ferromagnetic CuFeSb. In addition to this imbalance in $N_{\sigma}(E_F)$, the average Fermi velocity $\langle v_F \rangle_{\sigma}$ contribute to $P_t$ in the ballistic regime. We find $\langle v_F \rangle_{\uparrow}$ $>$ $\langle v_F \rangle_{\downarrow}$, and the calculated average Fermi velocity for the up and down channel is 6.44 $\times$ 10$^5$ and 2.68 $\times$ 10$^5$ m/s, respectively.  These result into a transport spin polarization, $P_t$ = 46\%, which is in excellent agreement with the present Andreev reflection spectroscopy measurement.

In conclusion, we have measured the transport spin polarization of ferromagnetic CuFeSb. From the analysis of the Andreev reflection spectra we obtain a spin polarization of approximately $47\%$. The first principles calculations show that the spin polarization in the Fermi surface is approximately $12.5\%$. However, when the transport spin polarization is calculated including the role of significantly different Fermi velocities for the up and down spin bands respectively, the ballistic transport spin polarization is found to be $46\%$ which is in excellent agreement with the experimental results. Furthermore, the band structure calculations shed light on the origin of a unique ferromagnetic ground state in CuFeSb as compared to the other compounds in the same family majority of which stabilize either in an antiferromagnetic or a superconducting ground state. 

We thank the employees running the liquid helium plant facility at IISER M that allowed us to perform the low-temperature measurements. We are grateful to Mr. Avtar Singh and Mr. Jithin Bhagwathi for their help during the experiments. MK $\&$ GS acknowledge partial financial support from the research grant of Ramanujan Fellowship awarded by the department of science and technology (DST), Govt. of India. AKG thanks DST, Govt. of India for financial assistance. GST and ZH thank
CSIR and UGC, Govt. of India for fellowship.

\end{document}